\documentclass[conference]{IEEEtran}

\ifCLASSINFOpdf

\else

\fi
\usepackage{tikz}

\newcommand\copyrighttext{%
  \footnotesize 2024 IEEE. Personal use of this material is permitted. Permission from IEEE must be obtained for all other uses, in any current or future media, including reprinting/republishing this material for advertising or promotional purposes, creating new collective works, for resale or redistribution to servers or lists, or reuse of any copyrighted component of this work in other works.}
\newcommand\copyrightnotice{%
\begin{tikzpicture}[remember picture,overlay]
\node[anchor=south,yshift=10pt] at (current page.south) {\fbox{\parbox{\dimexpr\textwidth-\fboxsep-\fboxrule\relax}{\copyrighttext}}};
\end{tikzpicture}%
}

\usepackage{booktabs}
\usepackage{amsmath,amssymb}
\usepackage{graphicx}
\usepackage{xspace}
\usepackage{color}
\usepackage{booktabs}
\usepackage{caption}
\usepackage{subcaption}
\usepackage{setspace}
\definecolor{azure}{rgb}{0.0, 0.5, 1.0}
\usepackage{pgfplots, tikzscale}
\usetikzlibrary{plotmarks}
\usetikzlibrary{plotmarks,patterns}
\pgfplotsset{compat=newest}
\usepackage{comment}
\usepackage[inline]{enumitem}

\graphicspath{ {figures/} }

\title{\fontsize{18.9}{18.9}\selectfont NL-COMM: Demonstrating Gains of Non-Linear Processing in Open-RAN Ecosystem}

\author{\IEEEauthorblockN{Chathura Jayawardena, Marcin Filo, George N. Katsaros and Konstantinos Nikitopoulos}
\IEEEauthorblockA{Wireless Systems Lab, 5G \& 6G Innovation Centre, \\Institute for Communication Systems, University of Surrey, Guildford GU2 7XH, UK}
}

\begin{document}

\maketitle
\IEEEpeerreviewmaketitle
\copyrightnotice
\begin{abstract}

Multi-user multiple-input, multiple-output (MU-MIMO) designs can substantially increase wireless systems' achievable throughput and connectivity capabilities.
However, existing MU-MIMO deployments typically utilize linear processing techniques that, despite their practical benefits, such as low computational complexity and easy integrability, can leave much of the available throughput and connectivity gains unexploited. They typically require many power-intensive antennas and RF chains to support a smaller number of MIMO streams, even when the transmitted information streams are of low rate. 
Alternatively, non-linear (NL) processing methods can maximize the capabilities of the MIMO channel. Despite their potential, traditional NL methods are challenged by high computational complexity and processing latency, making them impractical for real-time applications, especially in software-based systems envisioned for emerging Open Radio Access Networks (Open-RAN). Additionally, essential functionalities such as rate adaptation (RA) are currently unavailable for NL systems, limiting their practicality in real-world deployments. 
In this demo, we present the latest capabilities of our advanced NL processing framework (NL-COMM) in real-time and over-the-air, comparing them side-by-side with conventional linear processing. For the first time, NL-COMM not only meets the practical 5G-NR real-time latency requirements in pure software but also does so within a standard-compliant ecosystem. To achieve this, we significantly extended the NL-COMM algorithmic framework to support the first practical RA for NL processing. The demonstrated gains include enhanced connectivity by supporting four MIMO streams with a single base-station antenna, substantially increased throughput, and the ability to halve the number of base-station antennas without any performance loss to linear approaches.

\end{abstract}

\vspace{-5pt}
\section{Introduction}

%
Multi-User Multiple-Input, Multiple-Output (MU-MIMO) technology has been central to the evolution of 
the latest generations of
wireless networks, driving significant connectivity and throughput gains by multiplexing multiple data streams over the same time and frequency resources \cite{goldsmith_capacity_2003}.
In a MU-MIMO system, the achievable throughput and connectivity gains are intrinsically linked to the system's capability to accurately detect and decode mutually interfering information streams. Current, first-generation MIMO developments predominantly utilize linear processing techniques such as Minimum Mean Square Error (MMSE) and Zero-Forcing (ZF). Linear approaches transform the mutually interfering MIMO channel into multiple single-user channels. This allows for the direct application of traditional single-user processing techniques for detection and decoding and easy system integration due to the applicability of legacy radio resource management (RRM) approaches such as rate adaptation (RA). RA is essential for determining the transmission rate that maximizes achievable throughput in wireless networks and also enables efficient resource allocation and user scheduling. 

Despite these practical advantages, and as we show in this demo, linear processing methods substantially underutilize the corresponding MIMO channel, leaving unexploited throughput and connectivity gains. 
To compensate for that, linear systems employ a massive
number of power-intensive antennas and RF chains to support a much smaller number of concurrently transmitted streams 
\cite{nikitopoulos_massively_2022,nikitopoulos_geosphere_2014,nikitopoulos2024towards}. Typically, the employed number of antennas with linear processing is twice the number of supported MIMO streams \cite{nikitopoulos_massively_2022}, even when the combined transmission rates of the mutually interfering streams are significantly below the system's achievable capacity.
This severe underutilization of the MIMO channel translates to unnecessary increases in power consumption due to the excessive use of RF elements and reduced connectivity capabilities. It is worth noting here that massive connectivity is a key requirement for future network deployments such as vehicular \cite{katsaros_vehicular} and private networks \cite{nikitopoulos2024towards}.

Alternatively, non-linear (NL) approaches focus on jointly processing the mutually interfering MIMO streams, accounting for the MIMO channel characteristics, without degrading its inherent capabilities. Even though non-linear processing approaches \cite{softoutputSD,SFSD,nikitopoulos_geosphere_2014} promise substantial throughput and connectivity gains, their exponentially scaling computational complexity with the number of concurrently supported MIMO streams
and their weak parallelization properties often render them unsuitable for real-time realizations, especially in systems that are heavily software-based as emerging Open Radio Access Network (Open-RAN) designs envisage \cite{softiphy,azariah2022survey}.
In addition, fundamental RRM functionalities, such as RA for NL processing approaches, do not currently exist.
It is worth mentioning here that NL processing cannot exploit legacy approaches to perform essential RRM functionalities as RA, as it operates directly on the MIMO interfering channel. 
This results in a highly challenging joint optimization problem, in which the error propagation effects between users must be considered when predicting the transmission rate that maximizes throughput based on the channel and SNR. 
%
Furthermore, all existing evaluations of NL processing  \cite{softoutputSD,SFSD,jayawardena_g_multisphere_2020} employ the same modulation order for all concurrently transmitted users, which can limit achievable throughput and restrict their practical application, especially when large-scale fading is involved.





\begin{figure*}[!ht]
\centering
 \begin{subfigure}[b]{0.49\columnwidth}
        \centering
        \includegraphics[width=0.99\columnwidth]{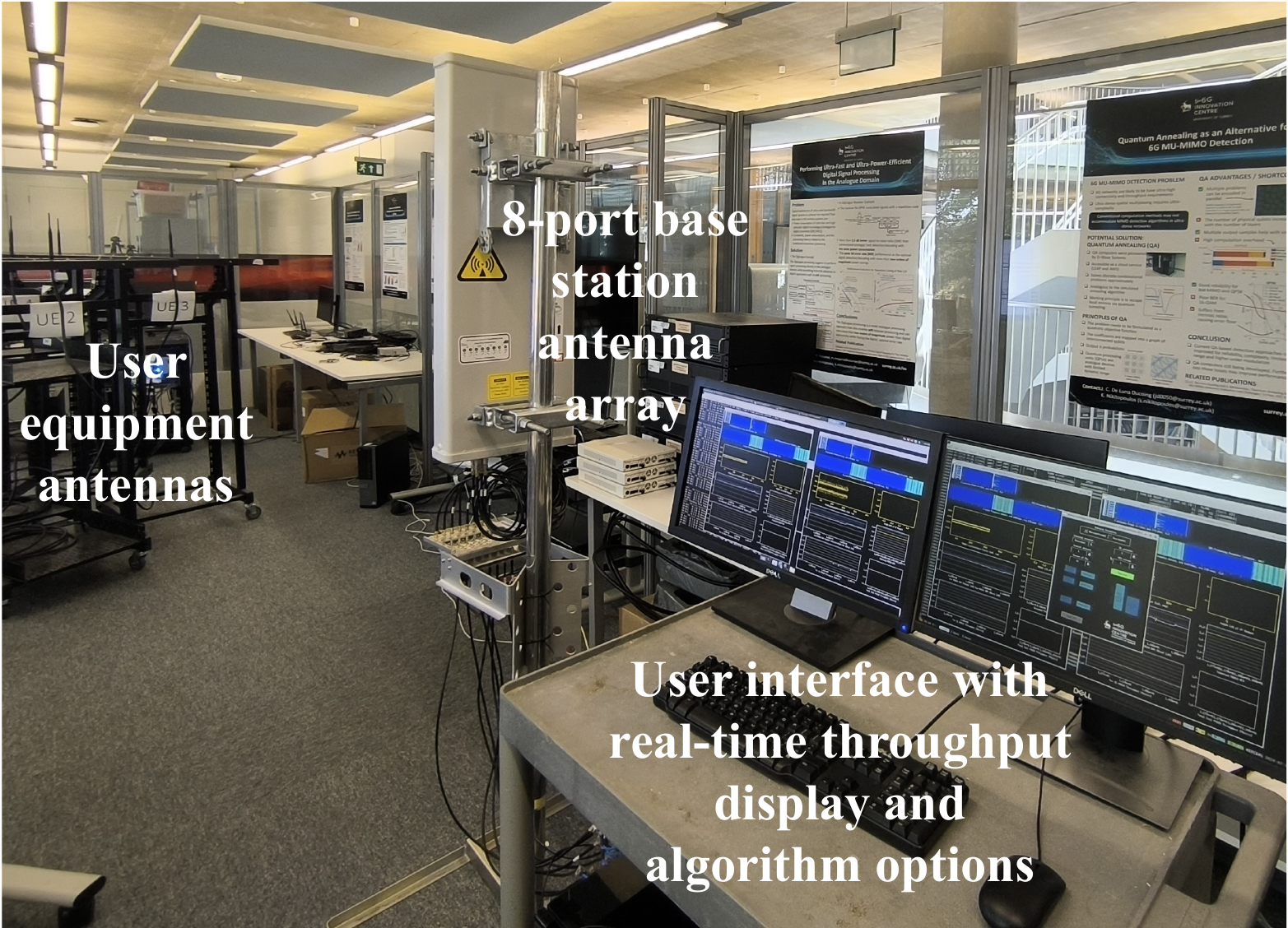}
        \caption{Demonstration Setup}
        \label{fig:DemoSetup}
    \end{subfigure}
    \begin{subfigure}[b]{0.49\columnwidth}
        \centering
        \includegraphics[width=0.99\columnwidth]{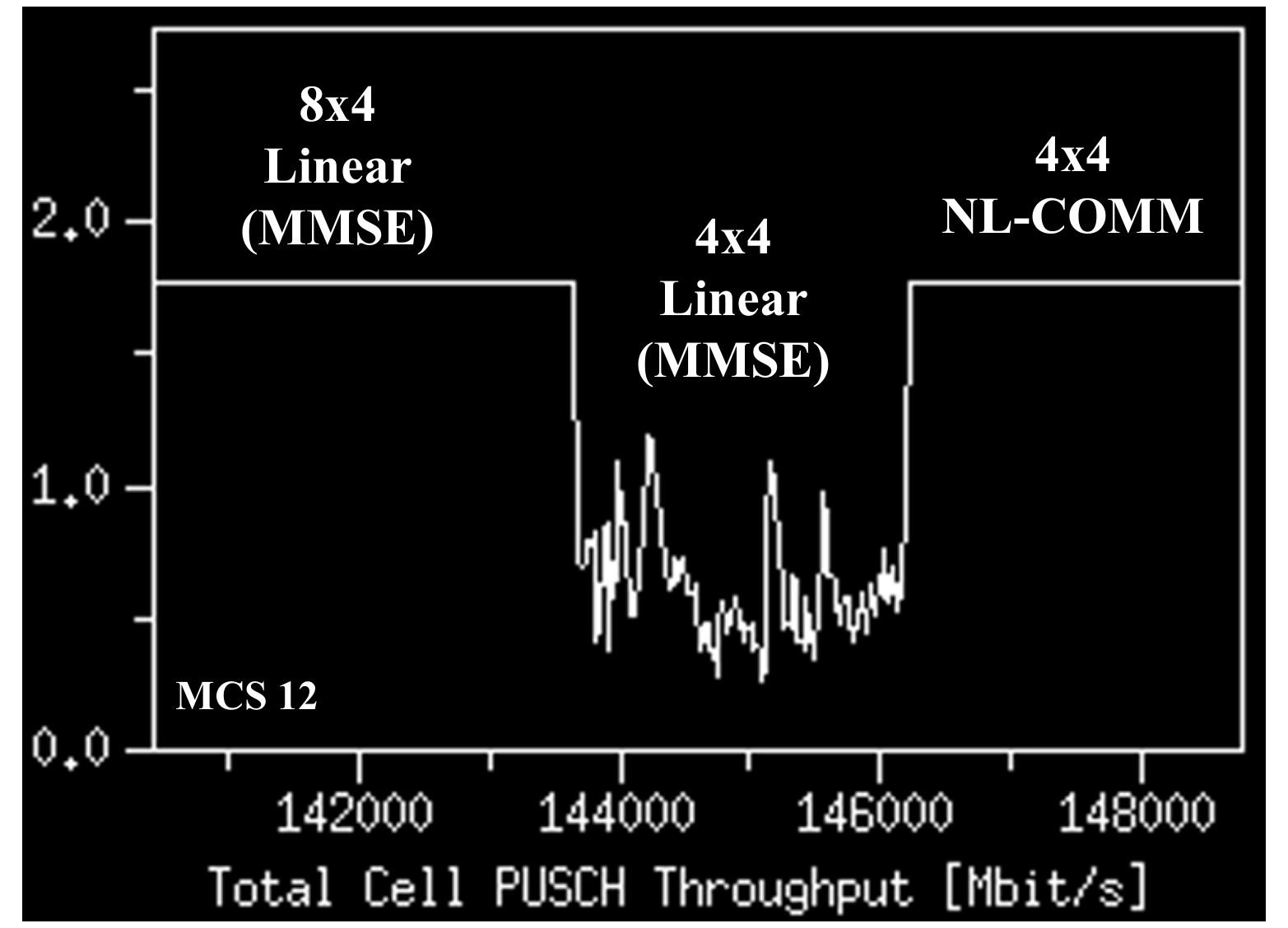}
        \caption{Halving of BS antennas}
        \label{fig:SimEv1}
    \end{subfigure}
    \begin{subfigure}[b]{0.49\columnwidth}
        \centering
        \includegraphics[width=0.99\columnwidth]{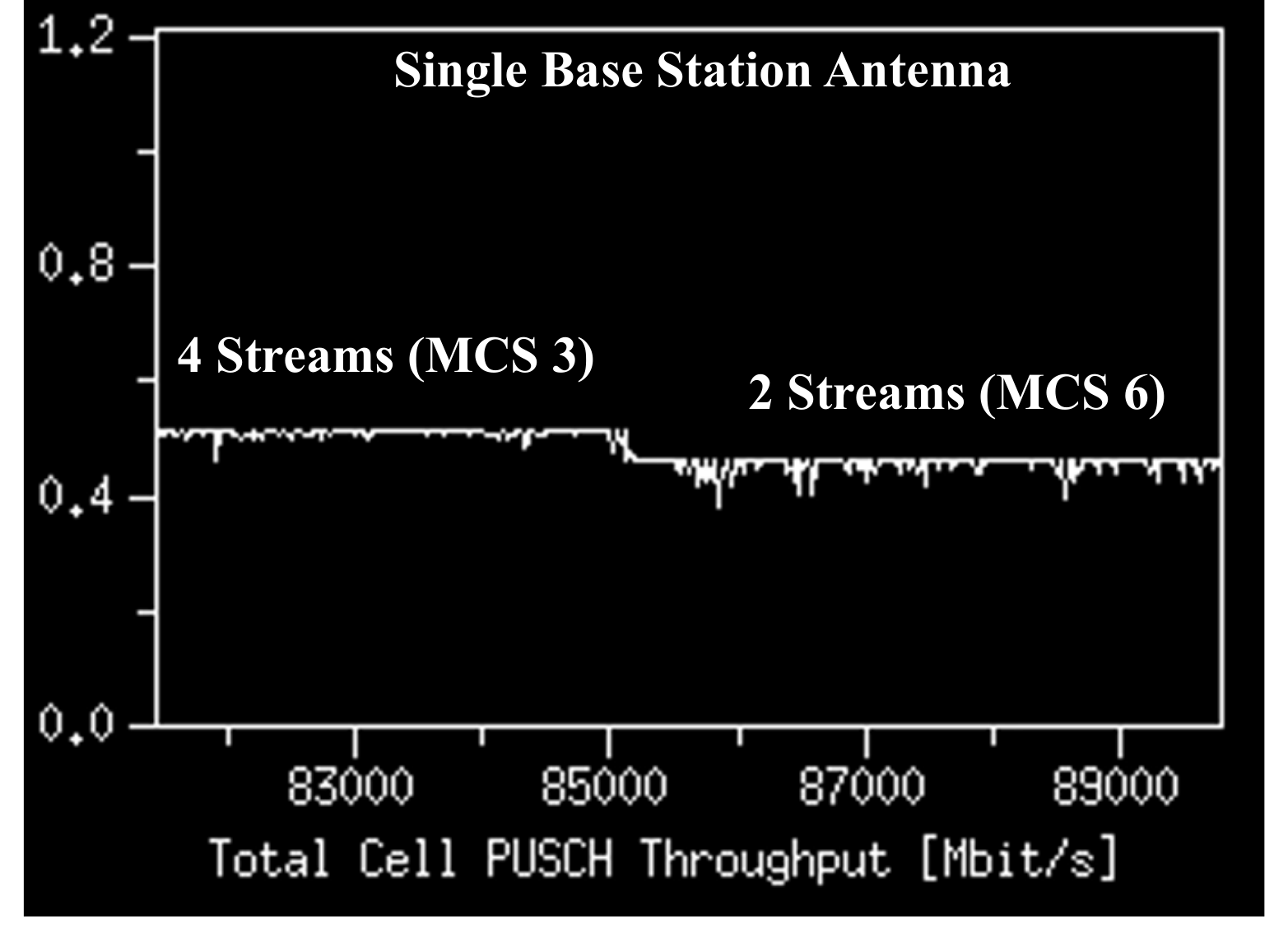}
        \caption{Enhanced connectivity}
        \label{fig:SimEv2}
    \end{subfigure}
    \begin{subfigure}[b]{0.49\columnwidth}
        \centering
        \includegraphics[width=0.99\columnwidth]{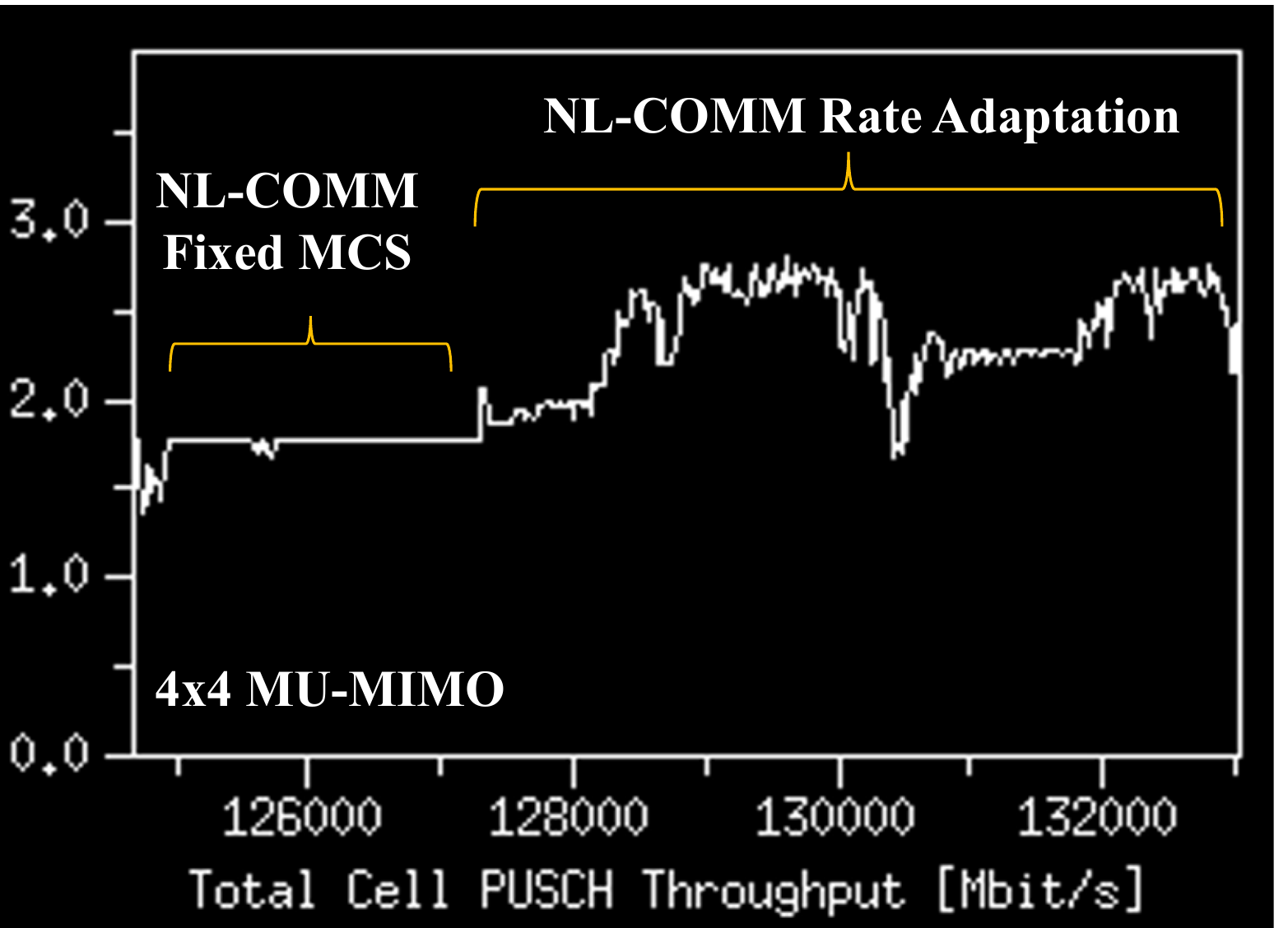}
        \caption{NL-COMM rate adaptation}
        \label{fig:SimEv3}
    \end{subfigure}
    \caption{\textbf{The demonstration setup (a) and indicative, real-time, and over-the-air throughput results ((b)-(d)).}}
    \vspace{-10pt}
     \label{fig:all}
\end{figure*}

In this demo, we present the latest capabilities of our advanced NL processing framework (NL-COMM). For the first time, NL-COMM not only meets practical 5G-NR real-time latency requirements in pure software but also does so within a 
fully standard-compliant ecosystem. 
To achieve this, we significantly extended the NL-COMM algorithmic framework to support the first practical RA for NL processing, without reliance on power-intensive machine learning approaches which struggle to adapt to rapidly varying wireless MIMO channel conditions.
Furthermore, this is the first real-time and over-the-air demonstration of NL-COMM's extreme connectivity capabilities, supporting four concurrently transmitting streams with a single base station antenna. This unlocks new possibilities for Non-Orthogonal Multiple Access (NOMA) schemes that typically require sparsity or changes to the standard to achieve smaller overloading factors.
The NL-COMM gains will be demonstrated and compared side-by-side with linear approaches. 
Indicative over-the-air results from our demonstration are shown in Fig. \ref{fig:all}.
The gains include, firstly, substantially increased throughput and the ability to halve the number of base-station (BS) antennas without any performance loss to linear processing approaches (Fig. \ref{fig:SimEv1}). Secondly, the ability to support a much larger number of users than BS antennas. Specifically, as shown in
Fig. \ref{fig:SimEv2} NL-COMM can efficiently overload more users than base-station antennas (400\%), improving channel utilization substantially. 
As shown, the channel's capacity can be shared among many users, while the lower the user's rate requirements, the higher the number of users that can be supported concurrently.
Finally, Fig. \ref{fig:SimEv3} shows the practical gains of NL-COMM for the first time with RA for NL processing, comparing it with a static Modulation and
Coding Scheme (MCS) scenario.

\section{Demonstration}

This demo utilizes our Open-RAN 5G-NR Stand Alone (SA) system, which supports MU-MIMO with linear and NL processing and is based on a substantially extended and optimized OpenAirInterface software platform.
Our system setup is depicted in Fig. \ref{fig:DemoSetup} and consists of a single 5G-NR base station with a commercial 8-port antenna array (hosted on a single DELL R740 server, and a single Ettus USRP X440 Software Defined Radio) connected to a 5G Core Network (hosted on a single DELL R740 server). 
The attendees will be able to change the placement of the UEs and, using a specially designed user interface, remotely alter various system parameters (e.g., RA algorithm type, MIMO detection algorithm type, MIMO streams) to validate the performance gains achieved by our RA algorithm. The impact of the changes on the system performance will be made available via a live stream transmission from our remote location. 
Our demonstration will require space for three 22-inch computer screens and high-speed internet access allowing for a live stream transmission.
An example preview of an earlier version of our demo can be found here:  \texttt{\textcolor{blue}{https://www.youtube.com/watch?v=0-c1OgnngDA}}
\section*{Acknowledgments}
This work has been supported by the ``NL-COMM'' \cite{nlcomm2024} project, a winner of the Small Business Research Initiative (SBRI) Competition by Innovate UK and UK’s Department for Science, Innovation and Technology (DSIT). Various aspects of the system and algorithm development have also been supported by the ``Flex5G'' \cite{flex5g2023} and ``HiPer-RAN'' \cite{hiper-ran2024} projects, winners of the UK’s DSIT Future RAN and Open Networks Ecosystem competitions.


\bibliographystyle{IEEEtran}
\bibliography{bib_demo_trunc}

\end{document}